\documentclass[runningheads]{llncs}
\usepackage[T1]{fontenc}
\usepackage{graphicx}
\usepackage{booktabs}
\usepackage{pifont}
\newcommand{\cmark}{\ding{51}}%
\newcommand{\xmark}{\ding{55}}%
\usepackage{amsmath}
\usepackage{amssymb}

\usepackage{multirow}
\usepackage{caption}
\usepackage[hidelinks]{hyperref}
\usepackage{cite} 
\usepackage{array}
\newcolumntype{x}[1]{>{\centering\arraybackslash\hspace{0pt}}m{#1}}
\newcolumntype{y}[1]{>{\raggedright\arraybackslash}m{#1}}

\def\fetmrqc{$\text{FetMRQC}_{\text{SR}}$}

\begin{document}
%
\title{Automatic quality control in multi-centric fetal brain MRI super-resolution reconstruction}
\titlerunning{Automatic QC for fetal brain MRI SRR}
%


\author{
Thomas  Sanchez\inst{1,2} \and
    Vladyslav Zalevskyi\inst{1,2} \and
     Angeline  Mihailov\inst{3} \and
     Gerard  Martí Juan\inst{4}\and
     Elisenda  Eixarch\inst{5, 6}\and
     Andras  Jakab\inst{7,8,9}\and
     Vincent  Dunet\inst{2}\and
     Mériam  Koob\inst{2}\and
     Guillaume  Auzias\inst{3}\and
     Meritxell  Bach Cuadra\inst{1,2}
}

\authorrunning{Thomas Sanchez et al.}

\institute{\email{thomas.sanchez@unil.ch}\\
\flushleft\footnotesize$^1$CIBM -- Center for Biomedical Imaging, Switzerland\\
\footnotesize$^2$Department of Diagnostic and Interventional Radiology, Lausanne University Hospital and University of Lausanne, Switzerland \\
\footnotesize$^3$Aix-Marseille Université, CNRS, Institut de Neurosciences de La Timone,\\\footnotesize Marseilles, France\\
\footnotesize$^4$BCN MedTech, Department of Engineering, Universitat Pompeu Fabra,\\\footnotesize Barcelona, Spain\\
\footnotesize$^5$BCNatal Fetal Medicine Research Center (Hospital Clínic and Hospital Sant Joan  de Déu), Universitat de Barcelona, Spain\\
\footnotesize$^{6}$IDIBAPS and CIBERER, Barcelona, Spain\\
\footnotesize$^7$Center for MR Research, University Children’s Hospital Zurich,\\\footnotesize University of Zurich, Switzerland\\
\footnotesize$^8$Neuroscience Center Zurich, University of Zurich, Switzerland\\
\footnotesize$^9$Research Priority Project Adaptive Brain Circuits in Development and Learning (AdaBD), University of Zürich, Switzerland
}

\maketitle              
\begin{abstract}
\vspace{-.5cm}
\looseness=-1
Quality control (QC) has long been considered essential to guarantee the reliability of neuroimaging studies. It is particularly important for fetal brain MRI, where acquisitions and image processing techniques are less standardized than in adult imaging. In this work, we focus on automated quality control of super-resolution reconstruction (SRR) volumes of fetal brain MRI, an important processing step where multiple stacks of thick 2D slices are registered together and combined to build a single, isotropic and artifact-free T2 weighted volume. We propose \fetmrqc, a machine-learning method that extracts more than 100 image quality metrics to predict image quality scores using a random forest model. This approach is well suited to a problem that is high dimensional, with highly heterogeneous data and small datasets. We validate \fetmrqc in an out-of-domain (OOD) setting and report high performance (ROC AUC = 0.89), even when faced with data from an unknown site or SRR method. We also investigate failure cases and show that they occur in $45\%$ of the images due to ambiguous configurations for which the rating from the expert is arguable. These results are encouraging and illustrate how a non deep learning-based method like \fetmrqc is well suited to this multifaceted problem. Our tool, along with all the code used to generate, train and evaluate the model are available at \url{https://github.com/Medical-Image-Analysis-Laboratory/fetmrqc_sr}

\end{abstract}

\section{Introduction}
\begin{figure}[!t]
    \vspace{-.2cm}
    \centering
    \includegraphics[width=\linewidth]{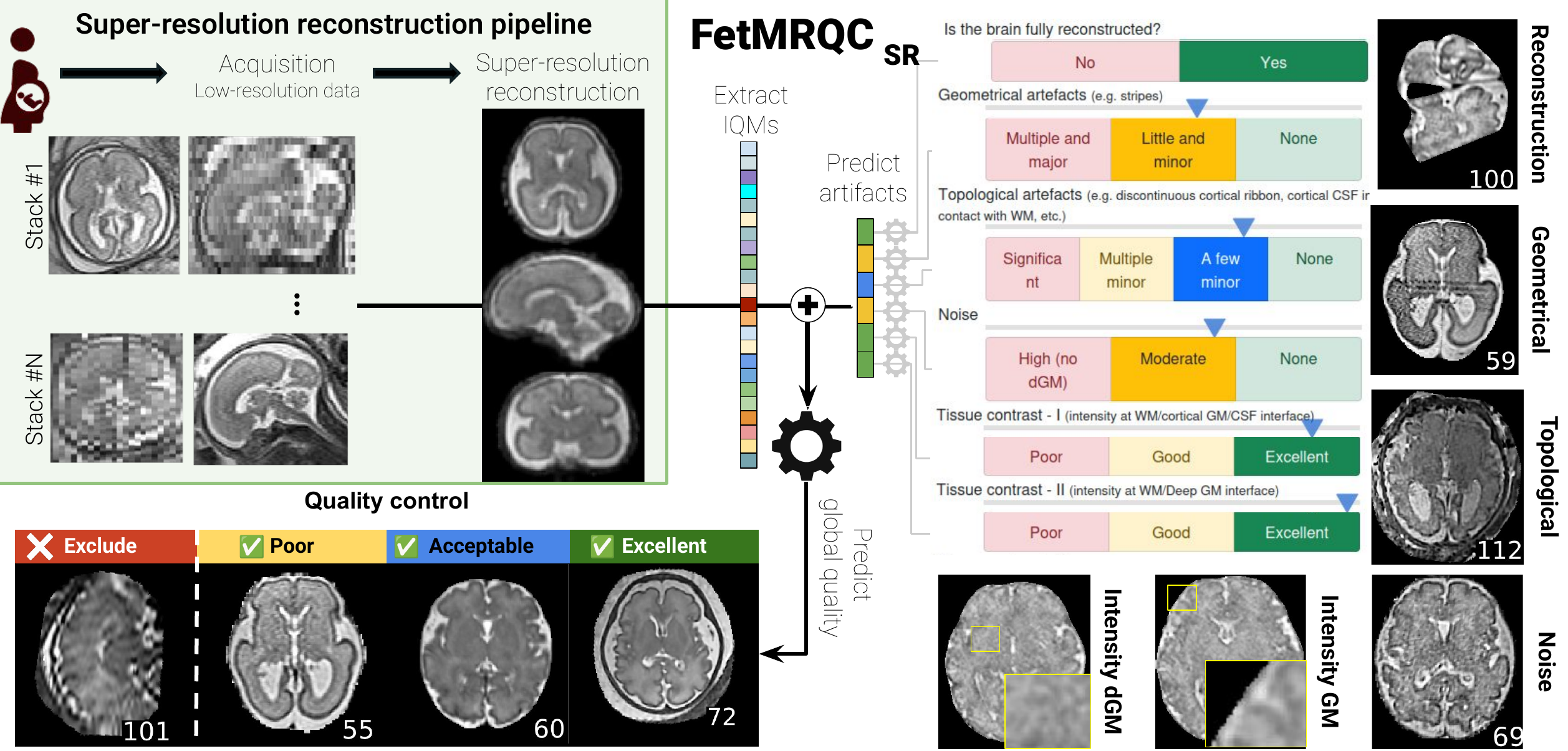}
    \caption{Illustration of \fetmrqc. \textbf{(Top left)} Several stacks of 2D T2-weighted slices are acquired and combined into a single super-resolution reconstructed (SRR) volume. \textbf{(Right)} SRR volumes are manually annotated according to several aspects, as well as with a global quality score \cite{sanchez2024assessing}. \textbf{(Bottom left)} \fetmrqc is trained to exclude SRR volumes below a certain quality threshold, based on image quality metrics (IQMs) extracted from the SRR volume.}
    \label{fig:qc_pipeline}
    \vspace{-.2cm}
\end{figure}

\looseness=-1
Quality control (QC) seeks to find and discard problematic outputs of a process to prevent problems from propagating~\cite{alfaro2018image}. This is fundamental in magnetic resonance imaging (MRI) studies, as insufficient MRI data quality has been shown to bias statistical analyses and neuroradiological interpretation~\cite{reuter2015head}. 
Image quality control protocols are thus increasingly seen as critical in achieving reliable, generalizable, and replicable results in neuroimaging studies~\cite{rosen2018quantitative,niso2022open}. 
Automated QC tools are popular for adult brain studies~\cite{esteban2017mriqc,klapwijk2019qoala,loizillon2024automated}. However, these techniques cannot easily be applied to fetal MRI, as they often rely on priors that are invalid \textit{in utero}, e.g. assuming that the head is surrounded by air, or rely on models of artifacts that differ from fetal MRI. Indeed, fetal brain MRI typically relies on the acquisition of multiple 2-dimensional (2D) fast spin-echo T2-weighted MR volumes with thick slices~\cite{tortori2005fetal,gholipour2014fetal}, followed by a post-processing step known as super-resolution reconstruction (SRR)~\cite{jiang2007mri,kuklisova-murgasova_reconstruction_2012,kainz_fast_2015,tourbier_efficient_2015,ebner_automated_2020,xu2023NeSVoR,uus2022retrospective}, where stacks of T2w slices are registered into a common space before being combined into a single, high resolution 3D volume (Figure~\ref{fig:qc_pipeline} -- top left). While these methods include built-in outlier rejection steps to increase their robustness, they can fail, returning a volume that will not be usable for further analysis (Figure~\ref{fig:qc_pipeline} -- bottom left). These volumes can feature various artifacts, such as topological inconsistencies, geometric distortions, high noise or low contrast between tissues~\cite{sanchez2024assessing} (Figure~\ref{fig:qc_pipeline} -- right).

\looseness=-1
Although the quality of SRR volumes greatly influences subsequent processing steps like automatic tissue segmentation and surface extraction, only a few methods have been proposed, and their validation was limited. Closest to this work, the study of Largent et al.~\cite{largent2021image} used multi-instance deep learning methods for automated quality control on 3D volumes reconstructed using only the approach from Jiang et al.~\cite{jiang2007mri}. The model trained in this work is not publicly available and the validation was  restricted to one acquisition site. Other works have taken a more descriptive approach to the quality of SRR volumes. The work of Rubert et al.~\cite{rubert2021data} focused on one popular SRR method called NiftyMIC~\cite{ebner_automated_2020}, and proposed to predict the quality of the reconstructed volume based on the percentage of slices from the input stacks that NiftyMIC uses for reconstructing that volume. The work of Sanchez et al.~\cite{sanchez2024assessing} proposed a protocol for fetal brain MRI quality assessment based on the evaluation of several image and artifact properties, alongside an overall image quality score~\cite{bach_cuadra_2025_15696638}. However, their sample size remained quite small and they did not perform an automated quality assessment. Finally, while a few automated QC methods have been proposed for low-resolution fetal brain MRI~\cite{xu2020semi,liao2020joint,sanchez2024fetmrqc}, they do not readily apply to SRR volumes, since the artifacts between raw T2-volumes and SRR volumes are very different. 

\looseness=-1
In this work, we propose \fetmrqc to tackle the challenge of automated QC for multi-SRR, multi-site data. Our contributions are as follows:
 \textbf{(i)}~We collect a unique dataset consisting of 673 manual quality ratings combining multiple acquisition sites and multiple SRR methods \textbf{(ii)}~We build upon FetMRQC~\cite{sanchez2024fetmrqc} to propose a robust approach based on random forest. We show that this non deep-learning approach, when combined with appropriate data pre-processing, provides a high performance, while requiring only a modest amount of training data. \textbf{(iii)}~We evaluate the out-of-domain robustness of our method, focusing on unseen sites and SRR methods. \textbf{(iv)}~We study the failure modes of our model and show that 45\% of the failures are due to ambiguous ratings, and that nearly all false negatives (wrongly flagged as bad quality) have some level of artifact that, although it is not disturbing to a human rater, is picked up by \fetmrqc. The code is publicly available on GitHub\footnote{\url{https://github.com/Medical-Image-Analysis-Laboratory/fetmrqc_sr}}. We believe that this method will serve as the first open source, easy-to-use baseline for future works in the field.


\section{Materials and methods}
\subsection{Demographics}
\looseness=-1
\noindent\textbf{Data.} Fetal brain MRI examinations were retrospectively collected from ongoing research studies at three hospitals: CHUV (n=141), BCNatal (n=180) and KISPI (n=37). The study received ethical approval from each center's institutional review board. These 358 subjects were acquired on different Siemens scanners (Erlangen, Germany) at 1.5T or 3T for CHUV and BCNatal, and on different GE Healthcare (Chicago, USA) scanners for KISPI.  As the aim of quality control is to be applicable to a wide range of scenarios, each subject was reconstructed using a randomly assigned set of parameters: a randomized SRR method, target resolution ($0.5^3/0.8^3/1.1^3\text{mm}^3$) and selection of input stacks. This resulted in 673 reconstructed volumes using four widely-used SRR methods: SVRTK~\cite{kuklisova-murgasova_reconstruction_2012,uus2022automated} (n=238),  NeSVoR~\cite{xu2023NeSVoR} (n=213), NiftyMIC~\cite{ebner_automated_2020} (n=202), MIALSRTK~\cite{tourbier_efficient_2015} (n=20). Additional details regarding the number of stacks used per reconstruction, echo and repetition times are provided in Table~\ref{tab:desc}.

\noindent\textbf{Manual quality ratings.}  After the SR reconstructions were completed, a manual quality assessment was performed following the protocol of Bach Cuadra et al.~\cite{bach_cuadra_2025_15696638}. The protocol was slightly adapted to focus on categories that were found to have an inter-rater reliability of $\text{ICC} > 0.5$, leading to the categories illustrated on the right of Figure~\ref{fig:qc_pipeline}. Global quality rating was obtained as a continuous variable between 0 and 4, and the exclusion threshold was set to 1. Two experts, M.B.C. (20 years of experience in fetal brain MR image analysis) and T.S. (4 years of experience) performed manual quality annotations on all the reconstructions (323 for M.B.C. and 448 for T.S., with 98 cases rated twice). The annotation interface allowed to streamline the process, and ratings took a total of 11h40 (7h10 for M.B.C., 4h30 for T.S.), with a median time of 47 seconds per case. Details on the distribution of SRR quality for the three main methods are provided in Figure~\ref{fig:qc}.

\subsection{Automated quality control}
Our automated quality control method, that we call \fetmrqc, is based on MRIQC~\cite{esteban2017mriqc} and FetMRQC~\cite{sanchez2024fetmrqc}. As illustrated on the right of Figure~\ref{fig:qc_pipeline}, \fetmrqc operates in two steps: 1)~extraction of a series of image quality measures (IQMs) from the input volumes, and 2)~training of a quality assessment model to predict the reference quality score, based on the IQMs. 

\noindent \textbf{Image quality metrics.} IQMs are features capturing complementary statistics on a given volume. Well-designed IQMs correlate with various image artifacts, enabling the training of a predictive QC model. However, IQMs that are commonly used in adult MRI are not readily applicable to fetal SRR volumes, as they can rely for instance on air as a background~\cite{esteban2017mriqc}, which is inappropriate \textit{in utero} as the fetus is surrounded by maternal tissues. This prompted us to adapt the IQMs of FetMRQC~\cite{sanchez2024fetmrqc}, designed for stacks of 2D slices of T2w fetal brain MRI, to the SRR setting. These IQMs are based on image intensity (e.g. mean intensity, SSIM of neighbouring slices, etc.), mask statistics (brain mask centroid coordinates) or segmentation-related statistics (tissue-wise statistics like volume, or contrast between tissue intensities). In addition to these metrics we incorporate new topological IQMs based on Betti numbers and Euler characteristics~\cite{hu2019topology}, as this has been shown to be an important metric to quantify the quality of a fetal brain MRI segmentation~\cite{payette2024multi,zalevskyi2025advances}.  
Since several IQMs required a brain mask and tissue segmentation, we used the state-of-the-art BOUNTI~\cite{uus2023bounti} to compute these maps. In total, 106 IQMs were computed (28 from FetMRQC, 54 from segmentation and 24 from topology). A detailed description of the IQMs used is provided in Table \ref{tab:iqms}.

\looseness=-1
\noindent \textbf{IQMs preprocessing.} 
Additionally we explore two pre-processing strategies to ensure reliable IQMs computation for all our data. In the first one, we saturate image intensities at 99.5 percentile prior to segmentation, as methods like NeSVoR~\cite{xu2023NeSVoR} occasionally produce extremely bright background voxels which in turn leads to a inaccurate segmentation. In the second, we apply brain masking, intensity saturation for low and high values ($0.05\to 0$ and $99.5\to 1$) and resizing to a 0.8mm resolution prior to the IQM computation.

\begin{figure}[t]
    \centering
    \includegraphics[width=0.4\linewidth]{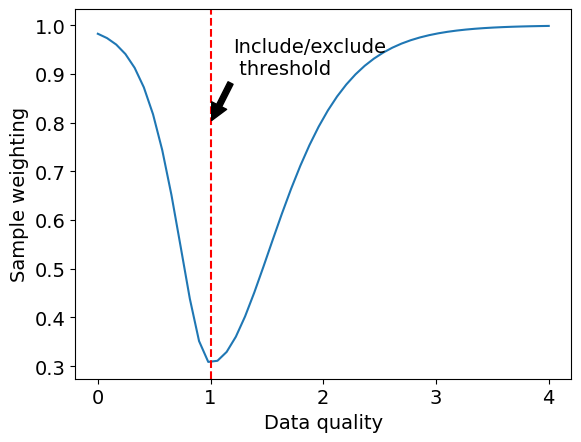}
    \caption{Sample re-weighting according to Equation~\ref{eq:weight}.}
    \label{fig:reweight}
\end{figure}
\noindent\textbf{Predictive model.}
\fetmrqc is based on a random forest, and trained in a binary classification setting, where it predicts whether a data sample has quality below the threshold of $1$, and should be excluded. In addition to it, we considered two variants of \fetmrqc. The first one was trained by re-weighted samples according to their quality score. Given a sample-label pair $(x,y)$, the continuous quality score $y$ ranging from 0 (exclude) to 4 (excellent) was mapped to a 0-1 probability $p(y)$ of being of good quality by using a sigmoid function centered at the decision boundary (quality score=1):
$$
p(y)=
\begin{cases}
\displaystyle 1/\left(1 + \exp\left(-6(y - 1)\right)\right) & \text{if } y < 1 \\
\displaystyle 1/\left(1 + \exp\left(-3(y - 1)\right)\right) & \text{otherwise}
\end{cases}.
$$
The sample $x$ then received a weight $w_x(y)$, based on its label, computed as the purity~($\triangleq$ 1 - the binary entropy $H$) of label $p(y)$:
\begin{equation}
    w_x(y) = 1 - H\big(p(y)\big) = 1 + p(y)\log p(y) + \big(1 - p(y)\big) \log\big(1 - p(y)\big).\label{eq:weight}
\end{equation}
We then have $w_x(0) = w_x(4) = 1 $, as these labels correspond to very unambiguous examples, and $w_x(1) = 0.3$ (minimal purity), as a sample at the decision boundary is inherently uncertain. It is referred to as \textsc{reweighting}, and shown in Figure~\ref{fig:reweight}. The second one, called \textsc{Predict\_artifacts}, made use of the available manual ratings of the type of artifacts present: random forest regressors were trained to predict each specific artifact using the IQMs (cf. Figure~\ref{fig:qc_pipeline} right), and then these predictions were used as additional IQMs for the \fetmrqc ~prediction. 

\noindent \textbf{Implementation.} \fetmrqc was implemented using \texttt{Python} 3.9.0, \texttt{scikit\-learn} 1.1.3, \texttt{Gudhi} 3.11.0 for topology computation and called on a docker wrapper of BOUNTI\footnote{\texttt{fetalsvrtk/segmentation:general\_auto\_amd}}. It builds on the code-base of FetMRQC~\cite{sanchez2024fetmrqc}\footnote{\url{https://github.com/Medical-Image-Analysis-Laboratory/fetmrqc}}.

\subsection{Experiment design}
\subsubsection{Evaluation setting.}
Previous studies have shown that domain shift is a major obstacle to the generalization of ML-based methods in medical imaging~\cite{dockes2021preventing}, motivating us to evaluate our models in four different experiments. \textbf{In-domain} (ID), where the data were split by subject across centers in 10-fold cross-validation (CV); \textbf{leave-one-site-out} CV, where all the data from a single site were used for testing; \textbf{leave-one-SRR-out} CV, where all the data from a given SRR were used for testing; \textbf{leave-one-site/SRR-out}, where all the data from a given site and a given SRR method were left out during training, and the model was tested on the combined data of the left-out site and SRR. This simulated a complex domain shift, where our method was tested on a new site, with data reconstructed using an unseen SRR method. 
Each experiment was repeated 10 times to account for random variations in the model training and training data sampling. All experiments used the same 673 samples but relied on different data splits. No pure testing evaluation was included in this paper due to space constraints.

\noindent\textbf{Metrics.} Our experimental setting being a classification task with imbalanced data, we reported the area under the ROC curve (AUC), balanced accuracy (BA), sensitivity, and specificity. Specificity is of particular interest, as we would like to especially avoid false positives, where bad quality data is wrongly predicted as being good. 

\looseness=-1
\noindent\textbf{Baselines.}
As no public baselines were available, we compared the following models and variants of \fetmrqc: \textbf{(i)~Stats.} A trivial lower-bound, statistics-based baseline, that predicts everything as good quality or bad quality based on the class frequency. \textbf{(ii)~\textsc{MLP}.} A model trained using a multi-layer perceptron (MLP) for prediction instead of a random forest.  \textbf{(iii)~\textsc{reweighting}.} A variant of \fetmrqc using sample re-weighting during training. \textbf{(iv)~\textsc{pred\_arti\-facts}.} A more complex variant of \fetmrqc, that first predicts the artifacts present in the image and then based on them predicts its quality score. \textbf{(v)~Oracle.} An oracle model that uses manual, reference ratings for the specific artifacts for the prediction of the global subjective score, serving as an upper bound on the performance of our method. 

\section{Results} 
\subsubsection*{Ablation study.}
\begin{figure}[!t]
\begin{minipage}[t]{0.35\linewidth}
\vspace{-.2cm}
\centering
\includegraphics[width=\linewidth,trim={5 15 7 5},clip]{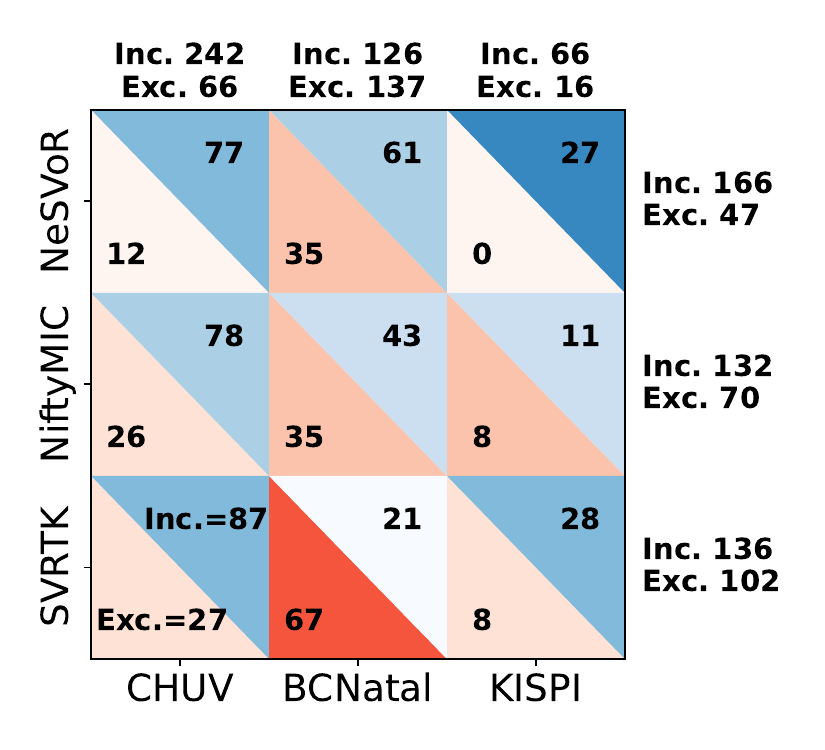}
\caption{Distribution of bad quality data (exclude; bottom in each cell, red) and good quality data (included; top, blue)  across sites and SRR method.}\label{fig:qc}
\end{minipage}
\begin{minipage}[t]{0.6\linewidth}
\vspace{0cm}
\centering

\captionof{table}{\textit{Ablation study.} For methods where not all thresholds are presented, only the threshold~$t$ leading to the highest BA was reported. Note that methods with different thresholds can still be compared at the ROC AUC level. Variance is computed over cross-validation folds.}
\label{tab:ablation}

\resizebox{\linewidth}{!}{
\begin{tabular}{lcccccc}
\toprule
    \multirow{2}{*}{\textbf{Model}} & \multicolumn{2}{c}{\textbf{Prepro.}}  &  \multicolumn{4}{c}{\textbf{In-domain generalization}} \\
    \cmidrule(rl){2-3}\cmidrule(rl){4-7}
    & Seg. & IQMs & AUC ($\uparrow$) & BA ($\uparrow$) & Sens. ($\uparrow$) & Spec. ($\uparrow$)\\ 
\midrule
    Stats & & & 0.50{\scriptsize$\pm$0.00} & 0.50{\scriptsize$\pm$0.00} & 0.36{\scriptsize$\pm$0.49} & 0.64{\scriptsize$\pm$0.49} \\
    MLP {\scriptsize(t=0.7)} & \cmark & \cmark& 0.87{\scriptsize$\pm$0.03} & 0.80{\scriptsize$\pm$0.04} & 0.85{\scriptsize$\pm$0.06} & 0.76{\scriptsize$\pm$0.11} \\
    \fetmrqc {\scriptsize(no. topo., t=0.7)} & \xmark & \xmark & 0.90{\scriptsize$\pm$0.03} & 0.80{\scriptsize$\pm$0.04} & 0.76{\scriptsize$\pm$0.11} & 0.84{\scriptsize$\pm$0.11} \\
    \fetmrqc {\scriptsize(no. topo., t=0.7)} & \cmark & \cmark & 0.90{\scriptsize$\pm$0.04} & 0.81{\scriptsize$\pm$0.04} & 0.78{\scriptsize$\pm$0.09} & 0.84{\scriptsize$\pm$0.11} \\
    \fetmrqc {\scriptsize(t=0.5)} & \xmark & \xmark & 0.91{\scriptsize$\pm$0.03} & 0.81{\scriptsize$\pm$0.07} & 0.91{\scriptsize$\pm$0.06} & 0.71{\scriptsize$\pm$0.19} \\
    \fetmrqc {\scriptsize(t=0.5}) & \xmark & \cmark & 0.91{\scriptsize$\pm$0.03} & 0.80{\scriptsize$\pm$0.05} & 0.91{\scriptsize$\pm$0.05} & 0.70{\scriptsize$\pm$0.13}  \\
    \fetmrqc {\scriptsize(t=0.5)} &  \cmark &  \xmark & 0.92{\scriptsize$\pm$0.02} & 0.81{\scriptsize$\pm$0.06} & 0.91{\scriptsize$\pm$0.06} & 0.72{\scriptsize$\pm$0.16} \\
    \fetmrqc {\scriptsize(t=0.5)} & \cmark & \cmark & \textbf{0.93{\scriptsize$\pm$0.02}} & 0.82{\scriptsize$\pm$0.06} & \textbf{0.92{\scriptsize$\pm$0.05}} & 0.72{\scriptsize$\pm$0.14} \\
    \textbf{\fetmrqc {\scriptsize(t=0.7)}} & \cmark & \cmark & \textbf{0.93{\scriptsize$\pm$0.02}} &\textbf{ 0.85{\scriptsize$\pm$0.03}} & 0.81{\scriptsize$\pm$0.08} &\textbf{ 0.88{\scriptsize$\pm$0.07}} \\
    \textsc{reweighting} {\scriptsize(t=0.7)} & \cmark & \cmark & \textbf{0.93{\scriptsize$\pm$0.02}} & 0.84{\scriptsize$\pm$0.02} & 0.81{\scriptsize$\pm$0.08} & \textbf{0.88{\scriptsize$\pm$0.08}}  \\
    \textsc{pred\_artifacts} {\scriptsize(t=0.5)}  & \cmark & \cmark &0.90{\scriptsize$\pm$0.03} & 0.81{\scriptsize$\pm$0.04} & 0.86{\scriptsize$\pm$0.08} & 0.76{\scriptsize$\pm$0.10} \\
    \midrule
    Oracle &  &  & 0.98{\scriptsize$\pm$0.01} & 0.92{\scriptsize$\pm$0.02} & 0.89{\scriptsize$\pm$0.05} & 0.95{\scriptsize$\pm$0.04}\\
\bottomrule
\end{tabular}}
\end{minipage}
\vspace{-.3cm}
\end{figure}

We first carried out an \textit{in-domain ablation} (subjects split using 5-fold CV) study on the various preprocessing options and variants of \fetmrqc. We also considered various thresholds $t$ for exclusion from the model's predicted probabilities, and investigated how changing $t$ impacted BA, sensitivity and specificity. Results are presented in Table~\ref{tab:ablation}, where we see that \fetmrqc using segmentation and IQMs pre-processing reaches the best performance with AUC=0.93 and BA=0.85. Choosing $t=0.7$ increases the specificity of the model, and we chose to keep it at that level for the rest of the paper. The addition of topological metrics improved the AUC and BA by 3\% and 4\% respectively. Finally, when looking at the oracle, which uses manual ratings of artifacts as the basis for prediction, we see that our models come close to the predictive ability of the oracle, but fall short by a margin of approximately 7-8\%. We also experimented with the site-SRR domain shift removal using ComBaT~\cite{johnson2007adjusting}, robust scaling, alternative classifiers like logistic regression, and nested cross-validation but these approaches did not improve the performance and are not reported. An analysis of the feature importances of \fetmrqc in a 10-fold CV setting is presented on Figure~\ref{fig:iqms}.

\begin{figure}[!t]
    \centering
    
    \vspace{0cm}
    \captionof{table}{\textit{Out-of-domain performance of the methods.} In the last case, the model does not use data from one particular site and one particular SRR at training time, simulating a double domain shift. }\label{tab:ood}
    \vspace{-.3cm}
    \resizebox{\linewidth}{!}{ \begin{tabular}{lcccccc|cccc|cccc}
    \toprule
    \multirow{2}{*}{\textbf{Model}} & \multicolumn{2}{c}{\textbf{Prepro.}} & \multicolumn{4}{c}{\textbf{Site}} & \multicolumn{4}{c}{\textbf{SRR}} & \multicolumn{4}{c}{\textbf{Site+SRR}} \\
    \cmidrule(rl){2-3}\cmidrule(rl){4-7}\cmidrule(rl){8-11}\cmidrule(rl){12-15}
    & Seg. & IQMs & AUC & BA & Sens. & Spec. & AUC & BA & Sens. & Spec. & AUC & BA & Sens. & Spec. \\ 
    \midrule
    Stats &  &  & 0.50{\scriptsize$\pm$0.00} & 0.50{\scriptsize$\pm$0.00} & 0.36{\scriptsize$\pm$0.49} & 0.64{\scriptsize$\pm$0.49} & 0.50{\scriptsize$\pm$0.00} & 0.50{\scriptsize$\pm$0.00} & 0.36{\scriptsize$\pm$0.49} & 0.64{\scriptsize$\pm$0.49} & 0.50{\scriptsize$\pm$0.00} & 0.44{\scriptsize$\pm$0.17} & 0.36{\scriptsize$\pm$0.49} & 0.53{\scriptsize$\pm$0.51} \\ 
    MLP (0.7) & \cmark & \cmark &0.85{\scriptsize$\pm$0.09} & 0.79{\scriptsize$\pm$0.09} & {0.82{\scriptsize$\pm$0.12}} & 0.75{\scriptsize$\pm$0.24} & 0.86{\scriptsize$\pm$0.05} & 0.79{\scriptsize$\pm$0.04} & {0.84{\scriptsize$\pm$0.09}} & 0.73{\scriptsize$\pm$0.13} & 0.83{\scriptsize$\pm$0.14} & 0.78{\scriptsize$\pm$0.10} & {0.82{\scriptsize$\pm$0.13}} & 0.66{\scriptsize$\pm$0.35} \\
    \cmidrule{1-15}
     & \xmark & \xmark & 0.90{\scriptsize$\pm$0.05} & 0.76{\scriptsize$\pm$0.09} & 0.65{\scriptsize$\pm$0.22} & 0.88{\scriptsize$\pm$0.13} & 0.86{\scriptsize$\pm$0.07} & 0.77{\scriptsize$\pm$0.07} & 0.63{\scriptsize$\pm$0.13} & 0.90{\scriptsize$\pm$0.11} & 0.85{\scriptsize$\pm$0.09} & 0.68{\scriptsize$\pm$0.11} & 0.53{\scriptsize$\pm$0.25} & \textbf{0.77{\scriptsize$\pm$0.34}} \\ 
     \textbf{\fetmrqc}& \xmark & \cmark & 0.90{\scriptsize$\pm$0.07} & 0.78{\scriptsize$\pm$0.09} & 0.68{\scriptsize$\pm$0.20} & \textbf{0.89{\scriptsize$\pm$0.10}} & 0.85{\scriptsize$\pm$0.06} & 0.76{\scriptsize$\pm$0.06} & 0.64{\scriptsize$\pm$0.15} & 0.88{\scriptsize$\pm$0.10} & 0.84{\scriptsize$\pm$0.08} & 0.73{\scriptsize$\pm$0.08} & 0.65{\scriptsize$\pm$0.22} & 0.74{\scriptsize$\pm$0.32} \\ 
     (t=0.7) & \cmark & \xmark & 0.92{\scriptsize$\pm$0.04} & 0.79{\scriptsize$\pm$0.09} & 0.69{\scriptsize$\pm$0.23} & \textbf{0.89{\scriptsize$\pm$0.06}} & 0.9{\scriptsize$\pm$0.04} & 0.8{\scriptsize$\pm$0.06} & 0.69{\scriptsize$\pm$0.15} & \textbf{0.91{\scriptsize$\pm$0.07}} & \textbf{0.89{\scriptsize$\pm$0.05}} & 0.73{\scriptsize$\pm$0.10} & 0.60{\scriptsize$\pm$0.27} & 0.77{\scriptsize$\pm$0.32}  \\ 
     & \cmark & \cmark & 0.92{\scriptsize$\pm$0.06} & 0.81{\scriptsize$\pm$0.09} & 0.74{\scriptsize$\pm$0.21} & \textbf{0.89{\scriptsize$\pm$0.08}} & \textbf{0.91{\scriptsize$\pm$0.06}} &\textbf{ 0.79{\scriptsize$\pm$0.07}} & 0.70{\scriptsize$\pm$0.14} & 0.88{\scriptsize$\pm$0.09} & 0.88{\scriptsize$\pm$0.05} & 0.79{\scriptsize$\pm$0.09} & 0.72{\scriptsize$\pm$0.22} & 0.75{\scriptsize$\pm$0.33} \\ 
     \cmidrule{1-15}
    \textsc{rew.} {\scriptsize(t=0.7)} & \cmark & \cmark & \textbf{0.93{\scriptsize$\pm$0.06}} &\textbf{ 0.82{\scriptsize$\pm$0.09}} & 0.75{\scriptsize$\pm$0.20} & \textbf{0.89{\scriptsize$\pm$0.07}} & 0.90{\scriptsize$\pm$0.06} & \textbf{0.79{\scriptsize$\pm$0.07}} & 0.69{\scriptsize$\pm$0.13} & 0.90{\scriptsize$\pm$0.08} & \textbf{0.89{\scriptsize$\pm$0.05}} & \textbf{0.80{\scriptsize$\pm$0.09}} & 0.72{\scriptsize$\pm$0.23} & \textbf{0.77{\scriptsize$\pm$0.33}} \\ 
    \midrule
    Oracle & ~ & ~ & 0.98{\scriptsize$\pm$0.02} & 0.92{\scriptsize$\pm$0.05} & 0.88{\scriptsize$\pm$0.06} & 0.95{\scriptsize$\pm$0.05} & 0.97{\scriptsize$\pm$0.02} & 0.92{\scriptsize$\pm$0.04} & 0.89{\scriptsize$\pm$0.08} & 0.94{\scriptsize$\pm$0.09} & 0.97{\scriptsize$\pm$0.03} & 0.91{\scriptsize$\pm$0.05} & 0.89{\scriptsize$\pm$0.07} & 0.83{\scriptsize$\pm$0.32} \\ 
    \bottomrule
\end{tabular}}
\vspace{-.2cm}
\end{figure}

\begin{figure}[!t]
\vspace{-.2cm}
    \centering
        \includegraphics[width=.8\linewidth,trim={2 0 0 5},clip]{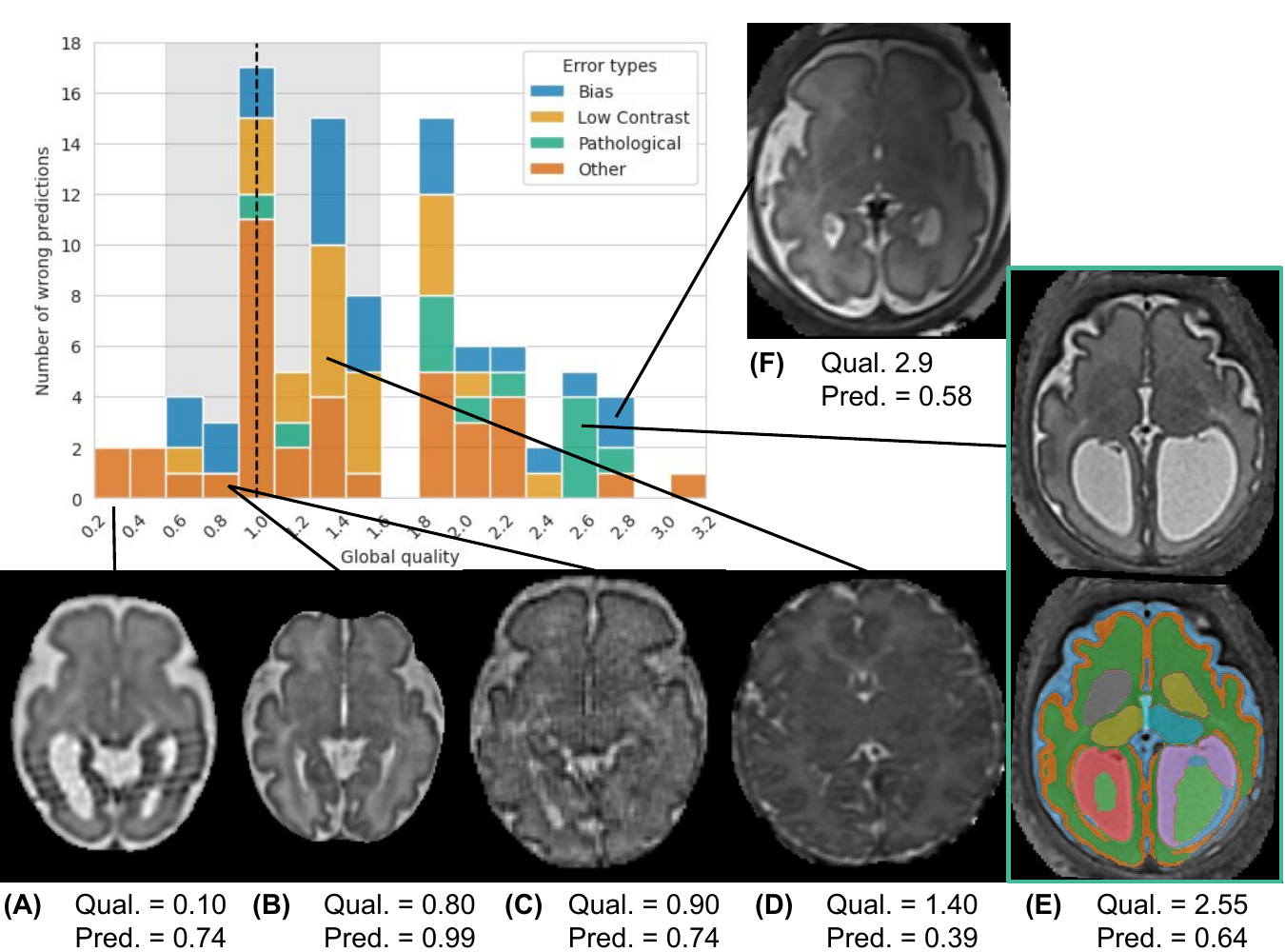}
    \vspace{-.2cm}
    \caption{\textit{Taxonomy of the failures of \fetmrqc}. The histogram shows the number of failed cases (95/673; 14\%) against the global quality, grouped into four main categories. The illustration shows different patterns, and reports the manual quality ratings (\textsc{Qual.}: threshold at 1.0 for exclusion) and the predicted probability of being excluded (\textsc{Pred.}: threshold $t$ at 0.7). \textbf{(A)} Substantial failure - geometric artifacts \textbf{(B)} Confident false positive - cut brain, \textbf{(C)} Borderline case -- \textsc{Qual.} and \textsc{Pred.} are close to their thresholds. \textbf{(D)} Confident false negative (FN) -- Low contrast. \textbf{(E)} Confident FN -- Pathological subject with failed segmentation. \textbf{(F)} Confident FN -- SRR with a heavy bias field.}
    \label{fig:failure}
    \vspace{-.7cm}
\end{figure}
\subsubsection*{Out-of-domain performance.}
\looseness=-1
In this experiment, we evaluate our model on \textit{three OOD settings}: leave-one-site-out, leave-one-SRR-out and leave-one-site/SRR-out CV. Our results, in Table~\ref{tab:ood}, show an increasing drop in performance across settings: it is less challenging to generalize to an unseen site than to an unseen SRR, and the most challenging setting is the double shift site+SRR. 
Although the AUC and BA do not vary much for \fetmrqc with robust preprocessing, there is a large drop in sensitivity when evaluating the model on a different site (-7\%) and SRR (-11\%) and a drop in both sensitivity and specificity in the site+SRR case (resp. -9\%; -13\%). 
The Site+SRR case also shows a larger variance in sensitivity and specificity of models, suggesting that the model might exhibit some degree of unreliability on this challenging domain shift. Nevertheless, \fetmrqc generally retains the best average performance. 

\subsubsection*{Failure modes.}
In this experiment, we assess cases where \fetmrqc fails. We perform 10 cross-validation experiments with random data splits and identify cases that are consistently mispredicted. We categorized the apparent causes of failure into four groups: bias field, low contrast images, pathological subjects and others. 

\looseness=-1
Out of 673 cases, the prediction was consistently false for 95 cases (14\%) across the repeated experiments. Figure~\ref{fig:failure} breaks down the distribution of the errors as a function of the manual quality rating. Out of the 95 failed cases, 23 correspond to cases with a strong bias field, and 22 to a low tissue contrast (Figure~\ref{fig:failure}D). Both were challenging for \fetmrqc's IQMs to capture and accurately classify. This is one limitation of feature engineering approaches that cannot capture all artifacts encountered. This is further illustrated by the cases with streaks and cut brain (Figure~\ref{fig:failure}A and B), that are confidently predicted as good quality, while being excluded by raters. Nevertheless, out of 95 failed cases, 42 had reference quality scores close to the decision boundary and we consider them as ambiguous cases, where a different rater could have assigned a different label (cf. Figure~\ref{fig:failure}C). These cases might be interpreted as limitations from the ground truth data as well as a challenging cases for \fetmrqc. Lastly, systematic failure cases also included images of obviously pathological cases, for which the automated segmentation method failed, leading in turn to inaccurate IQMs and mis-classification by the model.

\section{Discussion}
\looseness=-1
In this work, we took a look at the complete cycle of data and model in the problem of QC of fetal brain MRI SRR:  we did not operate only on improving our model given the data, but also investigated limitations within the data. Along with a rigorous validation of our model in a leave-one-site/SRR-out cross-validation setting, this enabled us to see that our classical, machine learning approach is well suited as a baseline for automated QC for fetal brain MRI SRR. Not only can it perform well even when data from a site and SRR are held out (with an AUC of 0.89), its failure modes can be attributed to the inability of IQMs to capture certain artifacts, as well as to the ambiguity in the data (44\% of failures are close to the decision boundary). QC is a multifaceted task, where different causes can lead to a similar rating, and annotations can vary across raters~\cite{sanchez2024assessing}, which translates into errors where it is not totally clear whether the problem lies in the prediction or the ground truth score. This variability in the data is further illustrated by the inability of the \textsc{Oracle} baseline to perfectly predict the global quality from specific artifact ratings. Moreover, as \fetmrqc depends on the robustness of its components, an inherent limitation originates from the segmentation method used, BOUNTI \cite{uus2022automated}, which is (as expected) unable to segment heavily pathological cases. This is where a deep learning-based approach could be promising: by learning correlations directly from data, it would not require a pre-trained segmentation model, and should capture more diverse artifacts than IQMs. We will investigate this option in future work and will use \fetmrqc as a baseline to evaluate the putative gain of a more complex approach. 

While not reported, we also experimented with CNN-based models\cite{wen2020convolutional}, which we found to be challenging this setting: training was very unstable, and we could not achieve stronger performance than \fetmrqc. We believe that DL-based models could be developed for this application, but found it to be not as easy as we would have expected and leave it for future work. Finally, we believe that including additional IQMs based on radiomics features~\cite{van2017computational} could also be an interesting avenue in the future.

Overall, this work highlights the importance of considering the entire machine learning development cycle, especially by looking at the data, and at failure cases~\cite{sambasivan2021everyone,zha2025data}, instead of only focusing on model development. Moreover, while the reported performance metrics are important, they fail to fully capture the behavior of a model, especially on subtle, multifaceted tasks like QC where ratings inherently contain some degree of uncertainty.

\begin{credits}

\vspace{-.2cm}
\subsubsection{\ackname}
This work was funded by Era-net NEURON MULTIFACT project (TS: Swiss National Science Foundation
grant 31NE30\_203977; AM, GA: French National Research Agency, Grant ANR-21-NEU2-0005; EE:
Instituto de Salud Carlos III (ISCIII) grant AC21\_2/00016 , GMJ: Ministry of Science,
Innovation and Universities: MCIN/AEI/10.13039/501100011033/), the SNSF project 215641, and the SulcalGRIDS Project,
(GA: French National Research Agency Grant ANR-19-CE45-0014). AJ is supported by the Prof. Max Cloetta Foundation, EMDO Foundation and Vontobel Foundation. This research was also supported by grants from NVIDIA through an Academic Grant Program and utilized the provided NVIDIA RTX6000 ADA GPUs. TS, VZ and MBC acknowledge access to the facilities and expertise of the CIBM Center for Biomedical Imaging, a Swiss research center of excellence founded and supported by CHUV, UNIL, EPFL, UNIGE and HUG. 
\vspace{-.2cm}
\subsubsection{\discintname}
The authors have no competing interests to declare that are relevant to the content of this article. 

\end{credits}
\newpage
%
%
%
\bibliographystyle{ieeetr}
{\small \bibliography{biblio}}
\newpage
\appendix
\makeatletter
\setcounter{table}{0}
\renewcommand 
\thesection{S\@arabic\c@section}
\renewcommand\thetable{S\@arabic\c@table}
\renewcommand \thefigure{S\@arabic\c@figure}
\makeatother
\section*{Supplementary material}

\begin{table}[h]
\centering
\caption{Expanded data description across sites, scanners and SRR methods. Number of SRR, numbers of stacks used for reconstruction, echo time (TE), repetition time (TR) and slice thickness of the acquired data are reported.}\label{tab:desc}
\resizebox{\linewidth}{!}{
\begin{tabular}{lllccccc}
\toprule
Site & Scanner & SRR & $n_{\text{SRR}}$ & $n_{\text{stacks}}$ & TE [s] & TR [s] & Thickness [mm] \\
\midrule
\multirow{9}{*}{\rotatebox{90}{BCNatal}} & \multirow{3}{*}{Aera (1.5T)} & NeSVoR & 35 & 11.17\text{\tiny±4.65} & 0.10\text{\tiny±0.03} & 1.27\text{\tiny±0.25} & 2.87\text{\tiny±0.44} \\
 &  & NiftyMIC & 20 & 8.10\text{\tiny±3.99} & 0.10\text{\tiny±0.02} & 1.18\text{\tiny±0.24} & 2.98\text{\tiny±0.41} \\
 &  & SVRTK & 20 & 9.25\text{\tiny±4.18} & 0.10\text{\tiny±0.02} & 1.22\text{\tiny±0.24} & 2.98\text{\tiny±0.44} \\
 \cmidrule[0.1pt]{3-8}

 & \multirow{3}{*}{TrioTim (3T)} & NeSVoR & 55 & 4.73\text{\tiny±1.96} & 0.14\text{\tiny±0.00} & 1.27\text{\tiny±0.43} & 3.83\text{\tiny±0.66} \\
 &  & NiftyMIC & 53 & 4.28\text{\tiny±1.50} & 0.14\text{\tiny±0.00} & 1.39\text{\tiny±0.73} & 3.76\text{\tiny±0.57} \\
 &  & SVRTK & 62 & 5.34\text{\tiny±3.21} & 0.14\text{\tiny±0.00} & 1.38\text{\tiny±0.74} & 3.74\text{\tiny±0.64} \\
\cmidrule[0.1pt]{3-8}
 & \multirow{3}{*}{Other} & NeSVoR & 7 & 4.00\text{\tiny±0.82} & 0.08\text{\tiny±0.01} & 1.42\text{\tiny±0.20} & 3.14\text{\tiny±0.24} \\
 &  & NiftyMIC & 5 & 4.20\text{\tiny±1.64} & 0.08\text{\tiny±0.00} & 1.55\text{\tiny±0.03} & 3.00\text{\tiny±0.00} \\
 &  & SVRTK & 6 & 5.00\text{\tiny±1.41} & 0.08\text{\tiny±0.00} & 1.51\text{\tiny±0.10} & 3.00\text{\tiny±0.00} \\

\midrule
\multirow{12}{*}{\rotatebox{90}{CHUV}} & \multirow{4}{*}{Aera (1.5T)} & MIALSRTK & 9 & 6.78\text{\tiny±2.05} & 0.13\text{\tiny±0.11} & 1.37\text{\tiny±0.51} & 2.86\text{\tiny±0.43} \\
 &  & NeSVoR & 47 & 7.68\text{\tiny±3.67} & 0.10\text{\tiny±0.07} & 1.26\text{\tiny±0.31} & 2.99\text{\tiny±0.29} \\
 &  & NiftyMIC & 62 & 6.61\text{\tiny±2.77} & 0.10\text{\tiny±0.06} & 1.25\text{\tiny±0.27} & 2.99\text{\tiny±0.30} \\
 &  & SVRTK & 60 & 8.12\text{\tiny±3.87} & 0.10\text{\tiny±0.06} & 1.25\text{\tiny±0.28} & 3.03\text{\tiny±0.35} \\
\cmidrule[0.1pt]{3-8}

 & \multirow{4}{*}{\begin{tabular}{l}
      MAGNETOM  \\ Sola (1.5T)
 \end{tabular}} & MIALSRTK & 8 & 6.50\text{\tiny±1.41} & 0.09\text{\tiny±0.00} & 1.20\text{\tiny±0.00} & 3.00\text{\tiny±0.00} \\
 &  & NeSVoR & 37 & 7.73\text{\tiny±3.82} & 0.10\text{\tiny±0.06} & 1.24\text{\tiny±0.25} & 2.98\text{\tiny±0.22} \\
 &  & NiftyMIC & 32 & 7.34\text{\tiny±3.83} & 0.09\text{\tiny±0.00} & 1.20\text{\tiny±0.00} & 3.00\text{\tiny±0.02} \\
 &  & SVRTK & 46 & 6.61\text{\tiny±2.44} & 0.11\text{\tiny±0.09} & 1.30\text{\tiny±0.38} & 2.93\text{\tiny±0.33} \\
\cmidrule[0.1pt]{3-8}

 & \multirow{4}{*}{Other} & MIALSRTK & 3 & 7.00\text{\tiny±1.73} & 0.09\text{\tiny±0.01} & 1.07\text{\tiny±0.06} & 3.33\text{\tiny±0.58} \\
 &  & NeSVoR & 5 & 7.00\text{\tiny±2.35} & 0.10\text{\tiny±0.00} & 1.10\text{\tiny±0.00} & 3.04\text{\tiny±0.09} \\
 &  & NiftyMIC & 11 & 6.73\text{\tiny±1.85} & 0.10\text{\tiny±0.01} & 1.09\text{\tiny±0.03} & 3.09\text{\tiny±0.30} \\
 &  & SVRTK & 8 & 8.62\text{\tiny±4.27} & 0.10\text{\tiny±0.00} & 1.10\text{\tiny±0.00} & 3.00\text{\tiny±0.00} \\
\midrule
\multirow{12}{*}{\rotatebox{90}{KISPI}} & \multirow{3}{*}{\begin{tabular}{l}
      DISCOVERY  \\ MR750 (3T)
 \end{tabular}} & NeSVoR & 2 & 4.50\text{\tiny±2.12} & 0.12\text{\tiny±0.00} & 3.00\text{\tiny±0.00} & 3.00\text{\tiny±0.00} \\
 &  & NiftyMIC & 5 & 8.80\text{\tiny±4.15} & 0.11\text{\tiny±0.01} & 2.32\text{\tiny±0.92} & 2.80\text{\tiny±0.27} \\
 &  & SVRTK & 5 & 14.80\text{\tiny±3.27} & 0.11\text{\tiny±0.01} & 2.45\text{\tiny±0.51} & 4.60\text{\tiny±1.52} \\
\cmidrule[0.1pt]{3-8}

 & \multirow{3}{*}{\begin{tabular}{l}
      DISCOVERY  \\ MR450 (1.5T)
 \end{tabular}} & NeSVoR & 6 & 7.67\text{\tiny±1.97} & 0.12\text{\tiny±0.00} & 3.00\text{\tiny±0.00} & 3.00\text{\tiny±0.00} \\
 &  & NiftyMIC & 2 & 8.50\text{\tiny±2.12} & 0.12\text{\tiny±0.00} & 3.00\text{\tiny±0.00} & 3.00\text{\tiny±0.00} \\
 &  & SVRTK & 5 & 7.20\text{\tiny±2.28} & 0.11\text{\tiny±0.01} & 2.68\text{\tiny±0.72} & 2.90\text{\tiny±0.22} \\
\cmidrule[0.1pt]{3-8}

 & \multirow{3}{*}{\begin{tabular}{l}
      SIGNA Artist  \\ (1.5T)
 \end{tabular}} & NeSVoR & 13 & 13.38\text{\tiny±6.38} & 0.12\text{\tiny±0.01} & 2.49\text{\tiny±1.08} & 3.08\text{\tiny±0.61} \\
 &  & NiftyMIC & 9 & 13.33\text{\tiny±4.09} & 0.12\text{\tiny±0.01} & 2.04\text{\tiny±0.94} & 3.17\text{\tiny±0.71} \\
 &  & SVRTK & 24 & 15.62\text{\tiny±7.27} & 0.12\text{\tiny±0.01} & 2.46\text{\tiny±0.92} & 2.98\text{\tiny±0.48} \\
\cmidrule[0.1pt]{3-8}

 & \multirow{3}{*}{\begin{tabular}{l}
      SIGNA Premier  \\ (3T)
 \end{tabular}} & NeSVoR & 6 & 13.83\text{\tiny±4.36} & 0.11\text{\tiny±0.01} & 2.55\text{\tiny±0.50} & 4.00\text{\tiny±1.10} \\
 &  & NiftyMIC & 3 & 9.67\text{\tiny±2.52} & 0.11\text{\tiny±0.01} & 2.15\text{\tiny±0.81} & 3.50\text{\tiny±1.32} \\
 &  & SVRTK & 2 & 15.50\text{\tiny±7.78} & 0.11\text{\tiny±0.02} & 2.52\text{\tiny±0.67} & 4.00\text{\tiny±1.41} \\
\bottomrule
\end{tabular}}
\vspace{-3cm}
\end{table}

\begin{table}[ht]
\centering
\caption{Summary of Image Quality Metrics (IQMs) used in \fetmrqc}
\label{tab:iqms}
\resizebox{0.9\linewidth}{!}{
\begin{tabular}{y{.25\linewidth}ly{.65\linewidth}}
\toprule
\multicolumn{3}{l}{
\textsc{\textbf{Intensity-based IQMs}}}\\ 
\midrule
\textbf{IQMs}& 
\textbf{Count} & \textbf{Description} \\
\midrule
\textit{rank\_error} \cite{kainz_fast_2015} & 2 & Measure the compressibility of the image using a low-rank approximation. Two variants: \textit{absolute} and \textit{relative} rank error~\cite{kainz_fast_2015}. \\
\textit{slice-wise difference} \cite{kainz_fast_2015, kuklisova-murgasova_reconstruction_2012,ebner_automated_2020} & 14  & Metrics for outlier rejection: normalized MAE, mutual information, cross-correlation, RMSE, PSNR, SSIM, joint entropy. Two variants to each metric: \textit{Mean}/\textit{Median} of differences across slices computed within 3 slice window (7 metrics $\times$ 2 variants)\\
\textit{entropy} \cite{esteban2017mriqc} & 1 & Entropy across the brain volume \\
\textit{sstats} \cite{esteban2017mriqc} & 7 & Mean, median, std, percentiles (5\%, 95\%), coefficient of variation, kurtosis on brain ROI \\
\textit{filter\_image} \cite{sanchez2024fetmrqc} & 2 & Sharpness estimation using \textit{Laplace} and \textit{Sobel} filters\\
\midrule
\multicolumn{3}{l}{
\textsc{\textbf{Masked-based IQMs}}}\\
\midrule
\textit{mask\_volume} \cite{sanchez2024fetmrqc} & 1 & Volume of the brain mask \\
\textit{centroid} \cite{dedumast2020translating} & 1 & Location of the brain mask centroid \\
\midrule
\multicolumn{3}{l}{
\textsc{\textbf{Segmentation-based IQMs}}}\\
\midrule
\textit{sstats} \cite{esteban2017mriqc} & 40 & Stats per region (WM, GM, CSF, BS, CBM): mean, median, 5th/95th percentile, kurtosis, std, MAD, voxel count (5 regions $\times$ 8 stats)\\
\textit{volume} \cite{esteban2017mriqc} & 5 & Volume of WM, GM, CSF, BS, CBM \\
\textit{SNR} \cite{dietrich2007measurement} & 6 & Signal-to-noise ratio per region and globally \\
CNR \cite{magnotta2006measurement} & 1 & Contrast-to-noise ratio between GM and WM \\
CJV \cite{ganzetti2016intensity} & 1 & Coefficient of joint variation of GM and WM \\
WM2MAX \cite{esteban2017mriqc} & 1 & Ratio of white matter to maximum intensity \\
\midrule
\multicolumn{3}{l}{
\textsc{\textbf{Topology-based IQMs}} (WM, GM, CSF, BS, CBM and whole brain)} \\
\midrule
Betti numbers BN0, BN1, BN2 \cite{hu2019topology} & 18  & Betti numbers describing the number of connected components (BN0), the number of loops or holes (BN1) and the number of cavities (BN2) for five brain regions (WM, GM, CSF, BS, CBM) and entire brain (3 Betti numbers $\times$ 6 regions)\\
Euler Characteristic \cite{hu2019topology} & 6 & Euler characteristic (= BN0 - BN1 + BN2) for each region and entire brain \\
\bottomrule
\end{tabular}}
    \centering
    \includegraphics[width=\linewidth, trim={0 0 0 20}, clip]{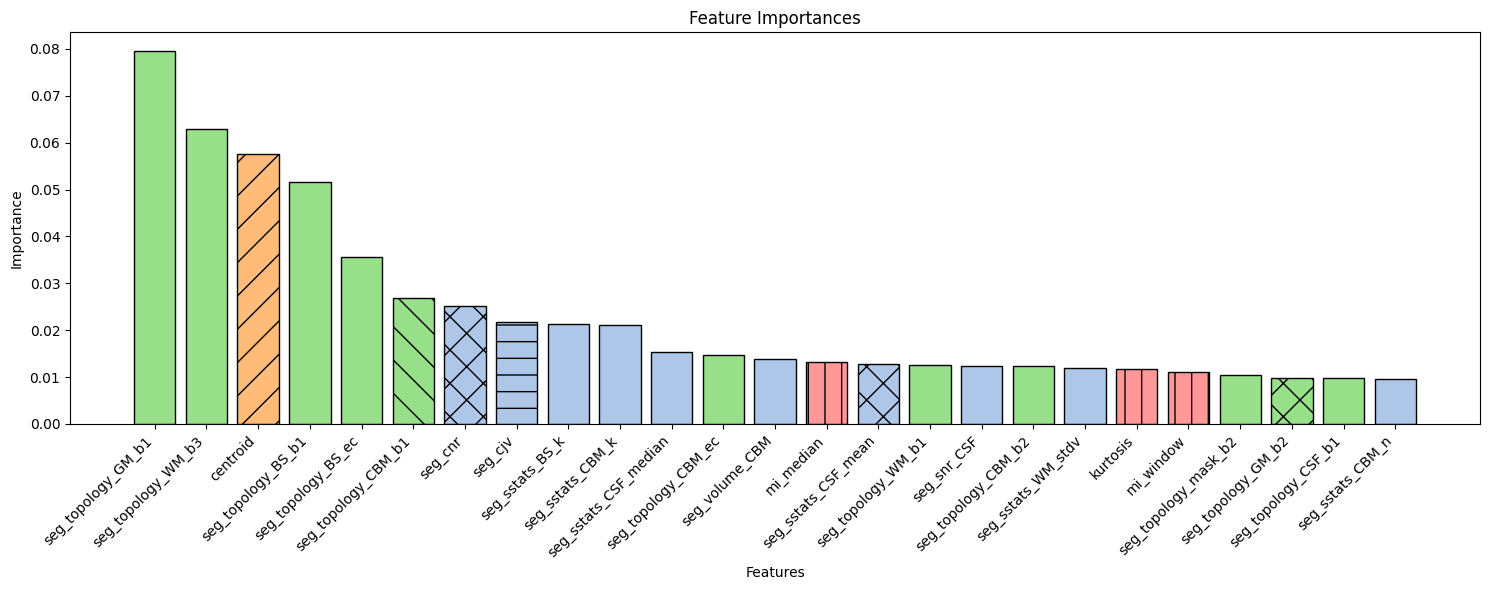}
    \captionof{figure}{Feature importances across 10 folds of cross validation for \fetmrqc. The results show the 25 most significant IQMs. Colors correspond the different categories (red: intensity, orange: mask, blue: segmentation, green: topology). Different hatching denote clusters  of highly correlated IQMs (Pearson correlation above $0.8$). Unhatched IQMs are not part of any cluster.}
    \label{fig:iqms}
    \vspace{-3cm}
\end{table}


\end{document}